\documentclass[fleqn,twoside]{article}
\usepackage{espcrc2}
\usepackage{amssymb}

\usepackage{graphicx}
\usepackage[figuresright]{rotating}

\newcommand{\AmS}{{\protect\the\textfont2
  A\kern-.1667em\lower.5ex\hbox{M}\kern-.125emS}}

\title{trans-GZK Cosmic Rays: Strings, Black Holes, Neutrinos, or all Three?}

\author{William S. Burgett\address[IfA]{Institute for Astronomy, 
        University of Hawaii \\ 
        2680 Woodlawn Dr. Honolulu, HI 96821}%
        \thanks{email: burgett@ifa.hawaii.edu}
        , Gabor Domokos\address[JHU]{Dept. of Physics \& Astronomy, Johns Hopkins University,\\
        $34^{th}$ and Charles St., Baltimore, MD  21218} 
        \thanks{email: skd@jhu.edu}
        , and Susan Kovesi-Domokos$^\mathrm{b\dag}$
        }
       
\begin{document}

\begin{abstract}
We review the scenario in which ``strongly interacting neutrinos'' are responsible
for inducing airshowers with inferred energies $E > 8\times10^{19}$ eV. This possibility
arises naturally in string excitation models having a unification scale effectively
decoupled from the Planck scale. We then show that phenomenological quantum gravity
considerations reveal an equivalency of ``mini-black hole'' and strongly interacting
neutrino pictures for explaining trans-GZK events. This equivalence can be exploited
to predict single particle inclusive distributions. The resulting observable
consequences in airshower development are studied using the Adaptive Longitudinal
Profile Shower (ALPS) simulation.
\vspace{1pc}
\end{abstract}

\maketitle

\section{INTRODUCTION}

The publication of the seminal papers by Greisen and Zatsepin \& Kuzmin~\cite{greisen}
predicting the energy attenuation of protons overs tens of megaparsecs 
due to interactions with CMB photons (GZK effect), together
with the observation of cosmic ray airshowers with inferred energies of
$\gtrsim 10^{20}$ eV, has created a vigorously pursued 
area of particle astrophysics.
In fact, as of 2004 there are many more
papers proposing explanations for the existence of
these trans-GZK airshowers than there are recorded events!
Due to this relatively sparse sample of events,
open questions include the existence of
statistically significant clustering/anisotropy, correlations
with known astrophysical source distributions (e.g., QSOs), 
the composition/charge of the primaries, and whether the production mechanism 
is top-down or bottom-up.
The present paper, delivered at the 2004 Cosmic Ray International Seminar (CRIS 2004), summarizes 
a bottom-up scenario whereby the primary incident on the Earth's atmosphere is
a neutral, non-hadronic particle such as a neutrino that generates a hadron-like
airshower due to the primary interaction being above a low-scale unification threshold.
Further details concerning the theoretical considerations sketched here, 
additional figures and tables, 
and more complete bibliographies can be found in the referenced papers.

\section{THEORETICAL MOTIVATIONS}

If trans-GZK events originate from sources farther away than $50 - 100$~Mpc, the
most likely candidate for the primary is some type of neutral particle. A Standard
Model neutrino can propagate over cosmological distances with little energy loss,
but, of course, the interaction cross-section $\sigma_{\nu N \to \ell X}$
is several orders of magnitude too low to generate the observed showers. However, it has
been conjectured for some time that perhaps a ``strongly interacting neutrino'' could exist 
due to some as yet undiscovered ``new physics'', and thus possess the required
interaction strength at the appropriate CM energies~\cite{berezinsky}. With the realization
that higher-dimensional string theories may allow interaction unification at energies many
orders below the standard 4-dimensional Planck scale of $\sim 10^{19} \mathrm{GeV}$~\cite{horava}, interest in the 
strongly interacting neutrino picture has been revived, and has resulted in new work involving
rigorous theoretical considerations within the framework of a specific model~\cite{domokos1}.
Additional features relative to cosmic ray airshowers have been extracted, although ongoing
work continues to refine the predictions relative to conventional airshower observables~\cite{domokos2}.

\subsection{Review of the String-inspired Scenario}

More specifically, if the CM energy of the neutrino-nucleon is above a ``low energy'' unification
scale, the neutrino-quark interaction is ``strong'', and a leptoquark resonant state can form,
$\sigma_{\nu N \to LQ \mathrm{res} \to \ell X}$. In establishing model properties using the phenomenology of
higher-dimensional string theories, it is worth noting that ``typical'' tree level amplitudes for
strings yield a cross section that is too small to account for the trans-GZK events~\cite{cornet}.
This does not rule out the present model, but does emphasize that weakly-coupled string theories
are inadequate here, and one must calculate with the strongly-coupled theory 
(and non-perturbatively when possible).
This is analogous to attempting to calculate quark interactions using QCD.

The basic building blocks of the model are unitarity of the S-matrix, a rapidly (exponentially) rising
level density of resonances, unification of interactions with strength $\sim$ Standard Model strong
interaction at string scale $M_{\ast}$, and duality of resonances in a given channel with Regge exchanges
in crossed channels. Express the total cross section in terms of
partial waves and absorption coefficients using the optical theorem,
\begin{eqnarray}
\sigma_{\nu q \to LQ \mathrm{res}} (s) = \frac{8\pi}{s}\sum_{j}^{N_0(s)} 
(2j + 1)(1 - \eta_j \cos(2\delta_j)) \mbox{.} \nonumber  
\end{eqnarray}
Note that as $s$ increases, the resonances are no longer purely elastic with absorption slowing the growth of the
total cross section, and where $N(s)$ is the resonance level and equals the maximum angular momentum; 
ignoring corrections $O(M_Z/M_\ast)$, $\sigma_{\nu q \to LQ\mathrm{res}}$ tends to a constant
value. The level density of resonances
is well represented for the first few resonances by 
\begin{eqnarray}
\qquad dN_0 \propto 1.24N_0 \approx \exp 1.24(s_0/M_{\ast}^2) \; \; \mbox{.} \nonumber 
\end{eqnarray}

\subsection{Strings and Quantum Black Holes}

A connection between string theories and quantum black holes is not a new idea, and there
is notable recent work in this area~\cite{amati}.
Extending this to the case of interest here, Domokos \& Kovesi-Domokos have demonstrated 
a non-perturbative equivalence of the string excitation model with that
of quantum ``mini black holes''~\cite{domokos3}.

The treatment in reference~\cite{domokos3} is based on a statistical mechanical analysis of the
previously introduced string model beginning with consideration of the microcanonical density matrix
of the final state,
\begin{eqnarray}
\qquad \rho = \sum_{\alpha} \vert N,\alpha \rangle \langle N,\alpha \vert \delta(e - NM_{\ast}) 
\; \; \mbox{.} \nonumber
\end{eqnarray}
A straightforward manipulation yields the standard passge to the canonical ensemble with well-defined
entropy $S$ and a temperature that asymptotically approaches the Hagedorn temperature,
$T_H = M_{\ast}/3\pi$. Recall that in modern QCD, the Hagedorn temperature $T_H$ is interpreted as marking the
deconfining phase transition from the low temperature hadron phase (quark confinement) to the
quark-gluon plasma. In terms of modern quantum gravity and string theory, the entropy
of a $d$-dimensional quantum black hole (QBH) equals the string entropy $S$ at $T_{QBH} \sim M_{\ast} \sim T_H$. 
It is straightforward to derive additional relations between $S$, $M_{\ast}$, $T_H$, and $R_{Schwarz}$. 

The asymptotic estimate then has important consequences in allowing a non-perturbative calculation of the single particle
inclusive distribution similar to the statistical mechanical analysis carried out at the tree level for specific string
models by Amati \& Russo (1999). 
Naturally, calculated quantities are a function of the number of ``extra''
dimensions in the model (i.e., the number of dimensions in addition to the conventional 4-dim space-time). 
Depending on model choices for characteristic scales, the resulting
multiplicities can be similar to those expected in reactions initiated by a heavy nucleus~\cite{domokos3}.

\subsection{The Transition from SM to Unified Regimes}

\begin{figure}
\includegraphics*[scale=0.4]{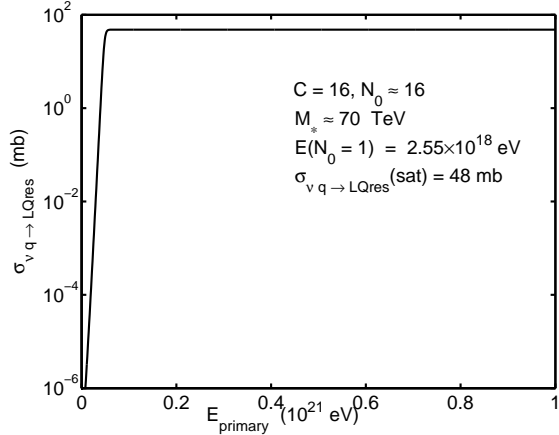}
\vspace*{-0.35in}
\caption{Smoothing function used for transition to unified interactions regime
showing the nearly step-function behavior due to the exponentially-rising level density.}
\label{smooth_fcn}
\end{figure}

\begin{figure}[t!]
\includegraphics*[scale=0.4]{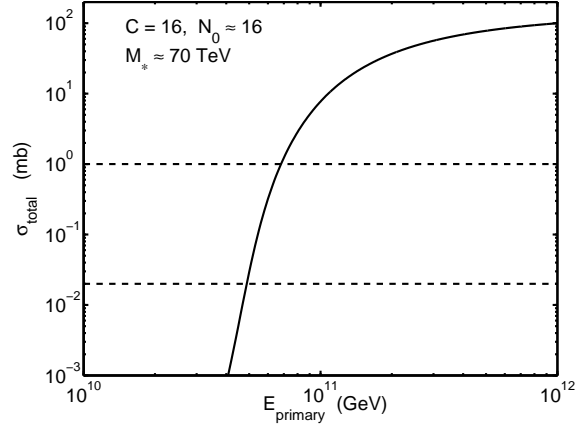}
\vspace*{-0.35in}
\caption{Integration of the approximately step-function neutrino-quark cross section over the
appropriate PDFs; the dashed lines indicate the approximate region that would generate observable deep showers.}
\label{int_pdf}
\end{figure}

Because the string-QBH model must contain resonances with the number of states an exponentially growing function
of the resonance mass, there is a (nearly) $\theta$-fcn transition from the SM to the unified regime. However,
there is obviously no exact $\theta$-fcn transition in Nature, thus implying that there must exist a finite width
transition region. With no analytic solution describing this transition available from the physics, we choose to
treat it empirically by adopting a convenient mathematical smoothing function to represent the neutrino quark
cross section. Nevertheless, the chosen form exhibits the desirable physical characteristics of nearly
step-function behavior saturating at a TBD strong interaction strength while remaining continuous and without violating
unitarity,
\begin{eqnarray}
\hat{\sigma}_{\nu q \to LQ\mathrm{res}}(s) = \mbox{~~~~~~~~~~~~~~~~~~~~~~~~~~~~~~~~~~~~~} \nonumber \\  
 \qquad \frac{16\pi}{M_{\ast}^2}\frac{C \exp 1.24 N_0} 
{1 + \frac{s}{M_{\ast}^2} \exp 1.24(N_0 - \frac{s}{M_{\ast}^2})}\theta(s - M_{\ast}^2) \mbox{.} \nonumber
\end{eqnarray}
In the above, the conversion from the Lab frame to the CM frame is via $s = 2m_{Nuc}E$, and $N_0 = 2m_{Nuc}E_0/M_\ast^2$
such that at $E = E_0$ we have $\hat{\sigma}_{\nu q \to LQ\mathrm{Res}}(E=E_0) = \frac{1}{2} \hat{\sigma}_{\nu q}(saturated)$.
The choices of $C$ and $N_0$ determine the value of $\hat{\sigma}_{\nu q} (saturated)$, taken here to be $\sim 50 - 100$~mb.
A typical example is shown in Figure~\ref{smooth_fcn} for a string scale $M_\ast \approx 70$ TeV.
Of course, the neutrino-quark cross section must be integrated over the parton distribution functions (PDFs)
of the participating partons in the nucleon to arrive at the final (observed) quark-nucleon cross section,
\begin{eqnarray}
\sigma_{\nu q \to LQ\mathrm{res} \to \ell X}(s,M_\ast,N_0) = \int_\frac{M_{\ast}^2}{s}^1 dx f(x) 
\hat{\sigma}(\hat{s}) \mbox{,} \nonumber
\end{eqnarray}
for $\hat{s} = xs$ and momentum fraction $x$. The integration is over valence quarks only since there
are no leptogluons in current string models, and the sea quarks contribute primarily around $x = 0$.
Figure~\ref{int_pdf} shows the result of integrating the $\hat{\sigma}_{\nu q \to LQ\mathrm{res}}$ 
of Figure~\ref{smooth_fcn} using the CTEQ6 PDFs. There is a region where the cross section would
give rise to deep showers, and this is indicated (approximately) by the dashed lines in the figure
(this is further discussed below).

\section{OBSERVABLE CONSEQUENCES}

\begin{figure}
\includegraphics*[scale=0.4]{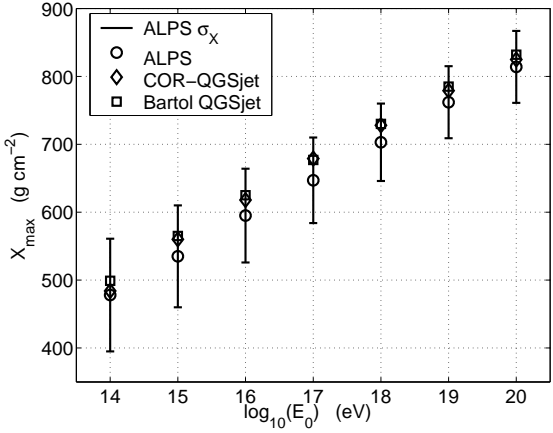}
\vspace*{-0.35in}
\caption{Comparison of $X_{max}$ values generated by ALPS, CORSIKA-QGSjet, and the Bartol-QGSjet
simulations for proton primaries. The error bars shown are from ALPS.}
\label{xmax_comp}
\end{figure}

Thus, as developed in the previous Section, the strength and properties of the unified interaction can lead to a
hadron-like airshower initiated by a specific neutral
string/QBH state (here taken to be a neutrino), but with some potentially
observable differences compared to those generated by hadrons or nuclei.
We study the generated airshower properties 
using the Adaptive Longitudinal Profile Shower (ALPS) simulation originally 
created by P. Mikulski~\cite{mikulski}. ALPS is similar to other hybrid simulations that use
subshower parameterization instead of tracking every individual particle. Its performance is
comparable to CORSIKA-QGSjet and the Bartol-QGSjet Hybrid Simulation, as shown in Figure~\ref{xmax_comp}
for average $X_{max}$ values generated by proton primaries as a function of primary energy.
Similar results hold for $N_{max}$ values.

\begin{figure}[t!]
\includegraphics*[scale=0.4]{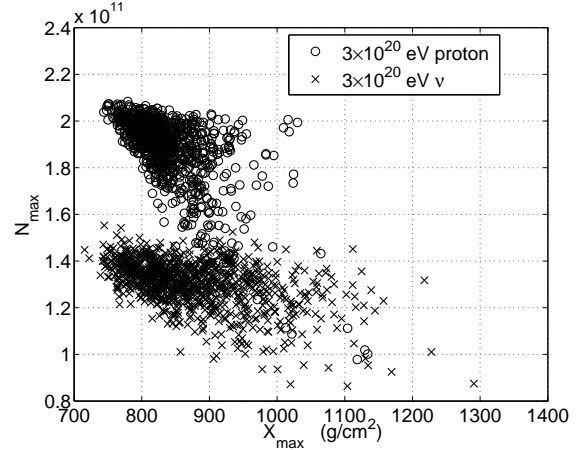}
\vspace*{-0.35in}
\caption{ALPS simulations of 1000 airshowers generated by $3\times10^{20}$ eV protons
and (strongly interacting) neutrinos.}
\label{alps_comp1}
\end{figure}
\begin{figure}[h!]
\includegraphics*[scale=0.4]{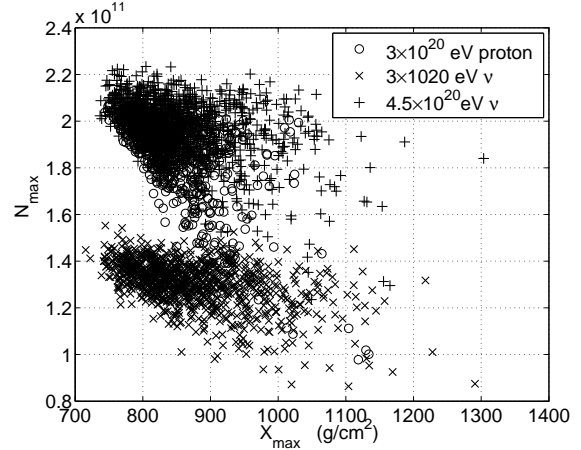}
\vspace*{-0.35in}
\caption{ALPS simulations of 1000 airshowers generated by $3\times10^{20}$ eV protons
and (strongly interacting) neutrinos and $4.5\times10^{20}$ eV neutrinos. The higher
energy neutrino events (`+' symbols) populate essentially the same region as the proton events
('o' symbols), and are not easily distinguished in this plot.}
\label{alps_comp2}
\end{figure}

Using the same string model parameters leading to Figures \ref{smooth_fcn} and \ref{int_pdf},
Figure~\ref{alps_comp1} presents ALPS results showing that, in general, a proton will
produce more electrons in a given airshower than a strongly interacting neutrino of the
same energy although the depth at maximum electron production is quite similar. The reason
for the $N_{max}$ difference is that a larger number of prompt leptons (mostly muons) are produced by the
neutrino. However, from Figure~\ref{alps_comp2}, it is also seen that a $4.5\times10^{20}$ eV
neutrino has both the approximately same $X_{max}$ and $N_{max}$ distributions as the
$3\times10^{20}$ eV proton, so that it would be virtually impossible to distinguish airshowers
generated by one or the other. In these cases, an additional airshower observable such as
muon number is required. Still, since it is expected that sources possess a non-monoenergetic
injection spectrum, over a large number of observed events with different primary energies,
the best discriminator may be identification of a specific source candidate since the source
distance can eliminate protons as candidate primaries. 
\begin{figure}
\includegraphics*[scale=0.4]{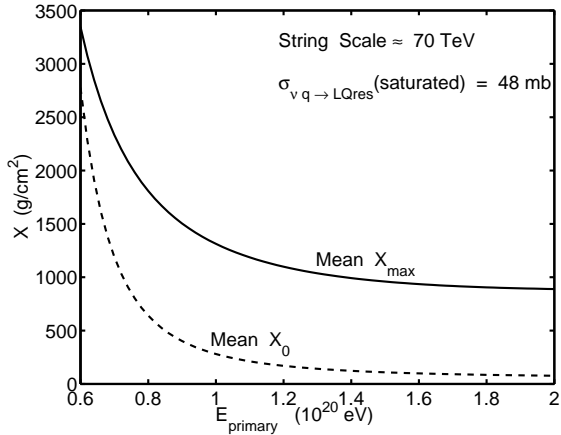}
\vspace*{-0.35in}
\caption{Preliminary extrapolations from ALPS simulations for the variation of $X_{max}$ and $X_0$
in the SM $\to$ unified transition region for a 70 TeV string scale.}
\label{alps_x}
\end{figure}

Another possible discrimination technique is the observation of highly inclined or
nearly horizontal airshowers since the expected width of the SM $\to$ unified transition
region should produce an excess of observed showers relative to that expected from
protons alone.  Predictions (extrapolations) for how $X_{max}$ and $X_0$ scale with
the changing cross section in the transition region are shown in Figure~\ref{alps_x}.
Also note that the fluctuations around the mean values shown in Figure~\ref{alps_x}
also increase very rapidly with decreasing cross section.
The interesting point here is that an observed flux of inclined showers
can be used to directly constrain the string scale, and, depending on the
model favored by Nature, this may be the \textit{only} way to do this in the next
decade (or beyond).

\section{DISCUSSION AND SUMMARY}

It should be mentioned that the idea that mini black holes might be produced in extra-dimensional theories
with energy scales orders of magnitude lower than the Planck scale has been
explored previously by several authors~\cite{feng}. However, the model presented
here differs in several important aspects:\\
\indent
{1. ``TeV-Scale Gravity is set by the SM electroweak scale (EW),} \\
\indent
{2. Interactions are not unified in such models (not a GUT), so there are no
strong scale interactions,} \\
\indent
{3. The cross sections for the black hole interactions are derived using a semi-classical
approach utilizing geometric total cross sections, $\sigma_{BH} \sim \pi R_{Schwarz}^2$,
and resonances are not included; while these cross sections are considerably enhanced compared
to SM EW values, they are still significantly smaller compared to SM strong cross sections.} \\

Currently, experimental results do not rule out the model discussed here, TeV-scale
gravity models, or many other competitors. However, the next generation detectors, especially
Auger and ASHRA, that are nearing completion are quite capable of discrminating among
the possibilities.

It is also interesting to note that a detailed analysis of the AGASA data appears to indicate that
the highest energy cosmic rays with $E > 8\times10^{19}$ eV may be distributed on the sky differently than
those having energies $4 < E < 8\times10^{19}$ eV, and this may support the idea that
the highest energy airshower events are generated by neutral particles that have propagated distances
in excess of 100 Mpc~\cite{burgett}. The main difficulty with such a scenario is that, at this
time, it is difficult to understand how a $\gtrsim 10^{20}$ eV neutrino is produced without first generating
a $\sim 10^{21} - 10^{22}$ eV hadron, and it is simply not known if such an astrophysical engine exists.
However, it is probably too soon to say with absolute certainty that such sources cannot exist.

Finally, we summarize the main points presented here. Our work is continuing in
both the theoretical and the simulation areas, and will be reported in future papers.
The robust features of the string/QBH phenomenology developed up to now include the
following:\\
\indent
{1. For given values of $C$ and $N_0$, the width of the SM $\to$ unified transition region
broadens as the characteristic string scale $M_\ast$ increases, independent of the choice
of smoothing function; a lower limit on the flux of deep showers corresponds to an upper
limit on $M_\ast$,} \\
\indent
{2. The energy at which the rising cross section reaches $1/2$ the saturated unified value determines
the presence (or lack) of a dip at high energies in the CR spectrum,} \\
\indent
{3. Exploiting the String-QBH equivalence provides a non-perturbative method for calculating
single particle inclusive distributions,} \\
\indent
{4. The presence of extra dimensions affects the single particle inclusive distributions in
a way that is potentially observable in cosmic ray airshowers, and may provide a mechanism
for determining the number of extra dimensions preferred by Nature.} \\
The model also yields a natural way to include an additional component in the EHECR distribution
if protons, pions, and neutrinos are produced in astrophysical engines with proton propagation
limited by the GZK effect, but with neutrinos able to propagate over cosmological distances.

\vspace*{0.25in}
\noindent{\textbf{Acknowledgments}}

The authors extend their appreciation and congratulations to the CRIS 2004 organizers and their
assistants for an informative, stimulating, and smoothly run conference.

\end{document}